# TRANS outperforms MTF for two special types of request sequences without locality of references

Rakesh Mohanty[1], Sangita Patel[2], Shiba Prasad Dash[3], BurleSharma[4]

[1]Department of Computer Science and Engineering, Indian Institute of Technology, Madras, Chennai, India -600036
[2]Dept. of Computer Sc. and Engg, Sambalpur University Institute of Information Technology, Burla, Odisha, India-768017.
[3, 4]Dept. of Computer Sc. and Engg, Veer Surendra Sai University of Technology, Burla, Sambalpur, Odisha, India-768018.

**Abstract**

Various list accessing algorithms have been proposed in the literature and their performances have been analyzed theoretically and experimentally. Move-To-Front (MTF) and Transpose (TRANS) are two well known primitive list accessing algorithms. MTF has been proved to be the best performing online algorithm till date in the literature for real life inputs and practical applications with locality of reference. It has been shown that when storage space is extremely limited and pointers for lists cannot be used, then array implementation of TRANS gives efficient reorganization. Use of MTF is extensive in the literature whereas, the use of TRANS is rare. As mentioned as an open problem in literature, direct bounds on the behavior and performance of various list accessing algorithms are needed to allow realistic comparisons. Since it has been shown that no single optimal permutation algorithm exists, it becomes necessary to characterize the circumstances that indicate the advantage in using a particular list accessing algorithm. Motivated by above challenging research issue, in this paper we have made an analytical study for evaluating the performance of TRANS list accessing algorithm using two special types of request sequences without locality of reference. We have compared the performance of TRANS with MTF and observed that TRANS outperforms MTF for these considered types of request sequences.

## 1. Introduction

Linear search is one of the basic search techniques for linear unsorted list. The efficiency of a liner search can be enhanced by making the list self organizing. List Update Problem (LUP) has been a popular problem for self organizing linear search. The input to the LUP is a list of distinct items, and a sequence of requests. Each request corresponds to an operation on an item of the list. The request may be either an access or insert or delete operation. Since insert and delete operations are special case of access operation, we consider only access operation for simplicity and hence the problem is also known as List Accessing Problem (LAP). When a request from a request sequence is served on the list, the requested item is accessed in the list by incurring some access cost using a cost model. After accessing the requested item, the list is reorganized so that the frequently accessed items are moved towards the front of the list to reduce the future access cost. In the LAP, our goal is to obtain the optimal access cost by efficiently reorganizing the list while serving a request sequence.

*1.1. List Accessing Cost Models*

The widely used cost models for LAP are full cost model and partial cost model. In full cost model the access cost of ith item in the list is i. Immediately after an access, the accessed item can be moved anywhere towards the front of the list without paying any cost. This type of exchange is called a free exchange. Any other exchange of two adjacent items in the list costs 1, and this type of exchange is known as paid exchange. In Partial cost model the cost of accessing the item is the number of comparisons made before accessing the item from the front of the list. The access cost of ith item in the list is (i-1), since it requires (i-1) comparisons before accessing the item i. The reorganization cost is the minimum number of paid exchanges. So the total cost is the sum of the access cost and the reorganization cost.

*1.2. List Accessing Algorithms*

An algorithm which reorganizes the list and minimizes the reorganization and access cost while serving a request sequence is called a list accessing algorithm. MTF, TRANS, and FC are the three basic primitive list accessing algorithm. In MTF, after accessing an item x in the list, x is immediately moved to the front of the list. In TRANS after accessing an item x in the list, x is moved forward one position in the list by exchanging it with the immediately preceding item. In FC a frequency counter is maintained for each of the items of the list as per the number of occurrences of each of the items in the request sequence. Whenever an item is accessed from the request sequence, the corresponding frequency counter is increased by one. The list is reorganized and maintained in non increasing order of the access frequencies at any instant of time.

*1.3. Applications and Motivation*

List accessing algorithms are extensively used for data compression. Some other popular applications of list accessing algorithms are maintaining small dictionaries, organizing the list of identifiers maintained by compilers and interpreters, resolving collisions in a hash table, computing point maxima and convex hulls in computational geometry.

The majority of the literature deals with analysis of various list accessing algorithms without any specific characterization of request sequences. Various patterns of request sequences occur in real life applications, out of which we have considered following two special types of request sequences as input. In the first type of request sequence, we have considered a permutation of the list which is in the same order of initial list configuration, being repeated more than once. In the second type of request sequence, we have considered a permutation of the list, which is in reverse order of the initial list configuration, being repeated more than once. The objective of our work is to evaluate the performance of MTF and TRANS algorithms using these two special types of request sequences and to compare the performance of both the algorithms.

*1.4. Literature review*

Study of list accessing problem was initiated by McCabe in 1965[1]. He proposed two popular algorithms MTF and TRANS. From 1965 to 1985, list accessing problem was studied by many researchers [2], [3], [4] with the assumption that the request sequence is generated by a probability distribution. Hester and Hirschberg [5] have provided an extensive survey of list accessing algorithms with some challenging open problems. Sleator and Tarjan [6] have shown the competitiveness of MTF using amortized analysis in their seminal paper. Reingold and Westbrook [8] have proposed an optimal off line algorithm for list accessing problem in 1996. Bachrach et.al.[7] have provided an extensive theoretical and experimental study of online list accessing algorithm in 2002. A study of list accessing problem with locality of reference was initiated by Angelopoulos in 2008[9]. A survey of important theoretical and experimental results related to on-line algorithms for list accessing problem is done in[10]. A classification of request sequences and few analytical results for MTF algorithm have been mentioned in [11].

*1.5 Our Contribution*

In our work we have considered two different types of request sequences corresponding to some real life inputs. Using these specific types of request sequences we have performed a theoretical and analytical study of MTF and TRANS list accessing algorithms and obtained some novel and interesting theoretical results. We have compared the performance of TRANS with MTF for these specific types of request sequences.

*1.6 Organization of the paper*

The paper is organized as follows. Introduction and literature review is presented in section I. Section II contains some novel analytical results for TRANS algorithm. Section III provides the concluding remarks and scope for future research work.

## 2 Novel analytical results for TRANS with special types of request sequences

In many real life applications the request sequence consists of one or more repetitions of different configurations of the list. In our work, we have considered some special types of request sequences that are repetitions of the same permutation of the list. Let $\ell = <\ell_1, \ell_2, \ell_3 .... \ell_n>$ be an unsorted list of $n$ items and $\sigma = <\sigma_1, \sigma_2, \sigma_3 .... \sigma_m>$ be a request sequence of size $m$ such that $\sigma_i \in \ell$ for i= 1, 2, 3,....m. For each item in the list, the list accessing algorithm must serve the request $\sigma_i$ in the order of its arrival. Let $k \geq 1$ be a positive integer that specifies the number of times a particular permutation of the list $\ell$ is repeated in the request sequence $\sigma$.

*2.1 Two special types of request sequences without locality of reference*

Online version of list accessing problem has become more significant in practical applications. An online algorithm knows only the current request that is to be served and the future request come in fly. Various online algorithms have been designed for the LAP and their performances have been analyzed by considering input request sequences with locality of reference. This locality of reference property suggests that the currently requested item is likely to be requested again in the near future. MTF has been proved to be the best performing online algorithm in the literature for request sequences with locality of reference. But for request sequences without locality of reference, the determination of best performing list accessing algorithm has not been done till date in the literature as per our knowledge. Here we have made an attempt to evaluate the performance of TRANS algorithm for two special types of request sequences without locality of reference. The above two types of request sequences are described as follows.

*Type1 ($T_1$)-* Let $\Pi_\ell = <\ell_1, \ell_2, \ell_3.... \ell_n>$ be a permutation of the list $\ell$ that consists of all the items of the list in the same order as in the list. $T_1 = \sigma = \Pi_\ell, \Pi_\ell ,.....$(k times) $= (\Pi_\ell)^k$ for k ≥1.

*Type2 ($T_2$)-* Let $\Pi_r = <\ell_n, \ell_{n-1}....\ell_2, \ell_1>$ be a permutation of the list $\ell$ that consists of all the items of the list in the reverse order of the list. $T_2 = \sigma = \Pi_r, \Pi_r ,.....$(k times) $=(\Pi_r)^k$ for k ≥1.

*2.2 Novel Analytical results for MTF*

Using request sequence of types T1 and T2, the following analytical results have been obtained for MTF algorithm. These results with proofs are presented in [12]. We present the results without proofs as follows for our reference. Here we consider the Full Cost Model and Singly Linked List as the data structure for our analysis.

**Theorem 1[12]-** *Let $C_{MTF}(\ell, T_1)$ be the total access cost incurred by MTF algorithm while serving a request sequence of Type $T_1$ on a list $\ell$ of size n then $C_{MTF}(\ell, T_1)=[n^2 \times (2k-1)+n]/2$ where $T_1 = (\Pi_\ell)^k$ for k ≥ 1.*

**Theorem 2[12]-** *Let $C_{MTF}(\ell, T_2)$ be the total access cost incurred by MTF algorithm while serving a request sequence $T_2$ on a list $\ell$ of size n then $C_{MTF}(\ell, T_2) = k \times n^2$.*

*2.3 Novel Analytical results for TRANS*

**Theorem 3.1-** *Let $C_{TRANS}(\ell, T_1)$ be the total access cost incurred by TRANS algorithm while serving a request sequence of Type $T_1$ on a list $\ell$ of size n, where $T_1 = (\Pi_\ell)^k$ for k ≥ 1, then*

a) $C_{TRANS}(\ell, T_1) = k \times \left[\frac{n^2+n+k-1}{2}\right]$     *when {k ≤ n/2 and n is even}*
    *or   {k ≤ (n-1)/2 and n is odd}*

b) $C_{TRANS}(\ell, T_1) = \left[\frac{n^2+2n}{2}\right] \times \left[k - \frac{1}{4}\right]$
    *when {k > n/2 and n is even}*

c) $C_{TRANS}(\ell, T_1) = \left[k \times \frac{n^2+2n-1}{2}\right] - \left[\frac{n^2-1}{8}\right]$
    *when {k > (n-1)/2 and n is odd}*

**Proof:** Let $C_{TRANS}(\ell, T_1)$ be the total access cost incurred by TRANS algorithm while serving a request sequence of type $T_1$ on a given list $\ell$ with initial list configuration $<\ell_1, \ell_2, \ell_3...., \ell_n>$. Let $T_{1i}$ be a subsequence of $T_1$ for *i=1, 2, 3...., k*. So $T_1 = T_{11}T_{12}T_{13}.... T_{1k}$ where each $T_{1i} = \Pi_\ell = <\ell_1, \ell_2, \ell_3.... \ell_n>$ for *i=1, 2, 3...., k*. Let $C_{TRANS}(L_i, T_{1i})$ be the total access cost of serving a request subsequence $T_{1i}$ of $T_1$ on a list configuration $L_i$. Here $L_i$ denotes a configuration of the list $\ell$ before serving the subsequence $T_{1i}$ for *i=1, 2, 3...., k*. The total access cost $C_{TRANS}(\ell, T_1)$ can be calculated as follows.
$C_{TRANS}(\ell, T_1) = \sum_{i=1}^{k} C_{TRANS}(L_i, T_{1i}) = C_{TRANS}(L_1, T_{11}) + C_{TRANS}(L_2, T_{12}) + .... C_{TRANS}(L_k, T_{1k})$.

**Step 1: Computation of $C_{TRANS}(L_i, T_{1i})$ for i=1**

Let $\sigma$ be a type $T_{11}$ request subsequence of $T_1$ that is served with list configuration $L_1 = \ell = <\ell_1, \ell_2, \ell_3.... \ell_n>$. Let $\sigma_j$ be the $j^{th}$ request of the request subsequence $\sigma$ and $C_{\sigma j}$(TRANS) denotes the access cost of serving a request $\sigma_j$ for j=1, 2, 3…., n using TRANS algorithm. So as shown in figure 1 $C_{\sigma1}$(TRANS) = 1, $C_{\sigma2}$ (TRANS) = 2, $C_{\sigma3}$ (TRANS) = 3….$C_{\sigma n}$ (TRANS) = n. Hence $C_{TRANS}(L_1, T_{11}) = \sum_{j=1}^{n} C_{\sigma j} (TRANS) = 1 + 2 + 3+….+ n = n(n+1)/2$ .

**Step 2: Computation of $C_{TRANS} (L_i , T_{1i})$ for i=2, 3,…, n/2 when *n* is even and i=2, 3,…, (n-1)/2 when *n* is odd**

$C_{TRANS} (L_i, T_{1i})= C_{TRANS} (L_1, T_{11})+(i-1)$ for each i=2, 3, 4…., n/2 , when n is even and for each i=2, 3, 4….,(n-1)/2, when n is odd. Hence $C_{TRANS}(L_2, T_{12})=(C_{TRANS}(L_1, T_{11})+1)$, $C_{TRANS}(L_3, T_{13})= (C_{TRANS}(L_1, T_{11})+2)$ and so on. Similarly $C_{TRANS}(L_{n/2}, T_{1n/2})= (C_{TRANS}(L_1, T_{11})+(n/2 - 1))$ when n is even and $C_{TRANS}(L_{(n-1)/2}, T_{1(n-1)/2})= (C_{TRANS}(L_1, T_{11})+[(n-1)/2 – 1]$ when n is odd.

**Step 3: Computation of $C_{TRANS} (L_i, T_{1i})$ for i > n/2 , when *n* is even and i > (n-1)/2 , when *n* is odd**

$C_{TRANS} (L_i, T_{1i})= C_{TRANS} (L_1, T_{11})+n/2$ for each i > n/2 when n is even and for each i > (n-1)/2 when n is odd.

**Step4: Computation of $C_{TRANS} (\ell, T_1)$**

**Proof of Theorem 3.1-a):** The complete illustration of TRANS algorithm for even value of n has been represented in figure 1. Let k ≤ n/2, when n is even and k ≤ (n-1)/2 when n is odd. Hence
$$C_{TRANS} (\ell, T_1) = C_{TRANS} (L_1, T_{11}) + \sum_{i=2}^{k} C_{TRANS}(L_i, T_{1i}) \quad (1)$$
From Step 2, $\sum_{i=2}^{k} C_{TRANS}(L_i, T_{1i}) = (C_{TRANS} (L_1, T_{11}) + 1) + (C_{TRANS} (L_1, T_{11}) + 2) +….[C_{TRANS}(L_1, T_{11}) + (k-1)] = (k-1) \times C_{TRANS}(L_1, T_{11}) + [1 + 2 + 3………+ (k-1)]$. So replacing the value of $\sum_{i=2}^{k} C_{TRANS}(L_i, T_{1i})$ in **(1)** we get $C_{TRANS} (\ell, T_1)= C_{TRANS} (L_1, T_{11})+(k-1)\times C_{TRANS} (L_1, T_{11}) +[1+2+3………+(k-1)] = k \times C_{TRANS} (L_1, T_{11}) + [1+ 2 + 3………+ (k-1)] = \left[k \times \frac{n(n+1)}{2}\right] + \left[\frac{(k-1)\times k}{2}\right] = k \times \left[\frac{n^2+n+k-1}{2}\right]$

**Proof of Theorem 3.1-b):** Let n is an even number and k > n/2. The total access cost $C_{TRANS}(\ell, T_1)$
$$=C_{TRANS} (L_1, T_{11})+ \sum_{i=2}^{n/2} C_{TRANS}(L_i, T_{1i}) + \sum_{i=\frac{n}{2}+1}^{k} C_{TRANS}(L_i, T_{1i}) \quad (2)$$
From step 2, $\sum_{i=2}^{n/2} C_{TRANS}(L_i, T_{1i}) = C_{TRANS}(L_2, T_{12})+ C_{TRANS}(L_3, T_{13})+… C_{TRANS}(L_{n/2}, T_{1n/2}) =C_{TRANS}(L_1, T_{11})+1)+(C_{TRANS}(L_1, T_{11})+2)+…[C_{TRANS}(L_1, T_{11})+(n/2 -1)]=(n/2-1) \times C_{TRANS}(L_1, T_{11}) +[1+2+3……+(n/2 - 1)] \quad (3)$

From step 3, $\sum_{i=\frac{n}{2}+1}^{k} C_{TRANS}(L_i, T_{1i}) =(C_{TRANS} (L_1, T_{11}) + n/2) + (C_{TRANS} (L_1, T_{11}) + n/2)…….(k - n/2)$ times $=(k - n/2) \times C_{TRANS}(L_1, T_{11}) +(k - n/2) \times n/2 \quad (4)$

Substituting the values of **(3)** and **(4)** in **(2)** we get, $C_{TRANS} (\ell, T_1)= C_{TRANS} (L_1, T_{11})+ (n/2 - 1)\times C_{TRANS}(L_1, T_{11})+ [1+2+3….+(n/2 - 1)]+ (k - n/2) \times C_{TRANS}(L_1, T_{11}) +(k - n/2) \times n/2 = k \times C_{TRANS}(L_1, T_{11})+ [1+2+3….+(n/2 - 1)]+ (k - n/2) \times n/2 = \left[k \times \frac{n(n+1)}{2}\right] + \left[\frac{(\frac{n}{2}-1)\times \frac{n}{2}}{2}\right] + \left[\left(k-\frac{n}{2}\right)\times \frac{n}{2}\right] = \left[k \times \left(\frac{n^2+2n}{2}\right)\right] + \frac{n^2-2n}{8} - \frac{n^2}{4} = \left[\frac{(n^2+2n)}{2}\right] \times \left[k - \frac{1}{4}\right]$

**Proof of Theorem 3.1-c):** Let n is an odd number and k > (n-1)/2. So the total access cost $C_{TRANS} (\ell, T_1)= C_{TRANS}(L_1, T_{11}) + \sum_{i=2}^{(n-1)/2} C_{TRANS}(L_i, T_{1i}) + \sum_{i=\frac{(n-1)}{2}+1}^{k} C_{TRANS}(L_i, T_{1i})$. So replacing the value of n/2 with (n-1)/2 in proof of b) we get, $C_{TRANS}(\ell, T_1) = \left[k \times \frac{(n^2+2n-1)}{2}\right] - \left[\frac{(n^2-1)}{8}\right]$

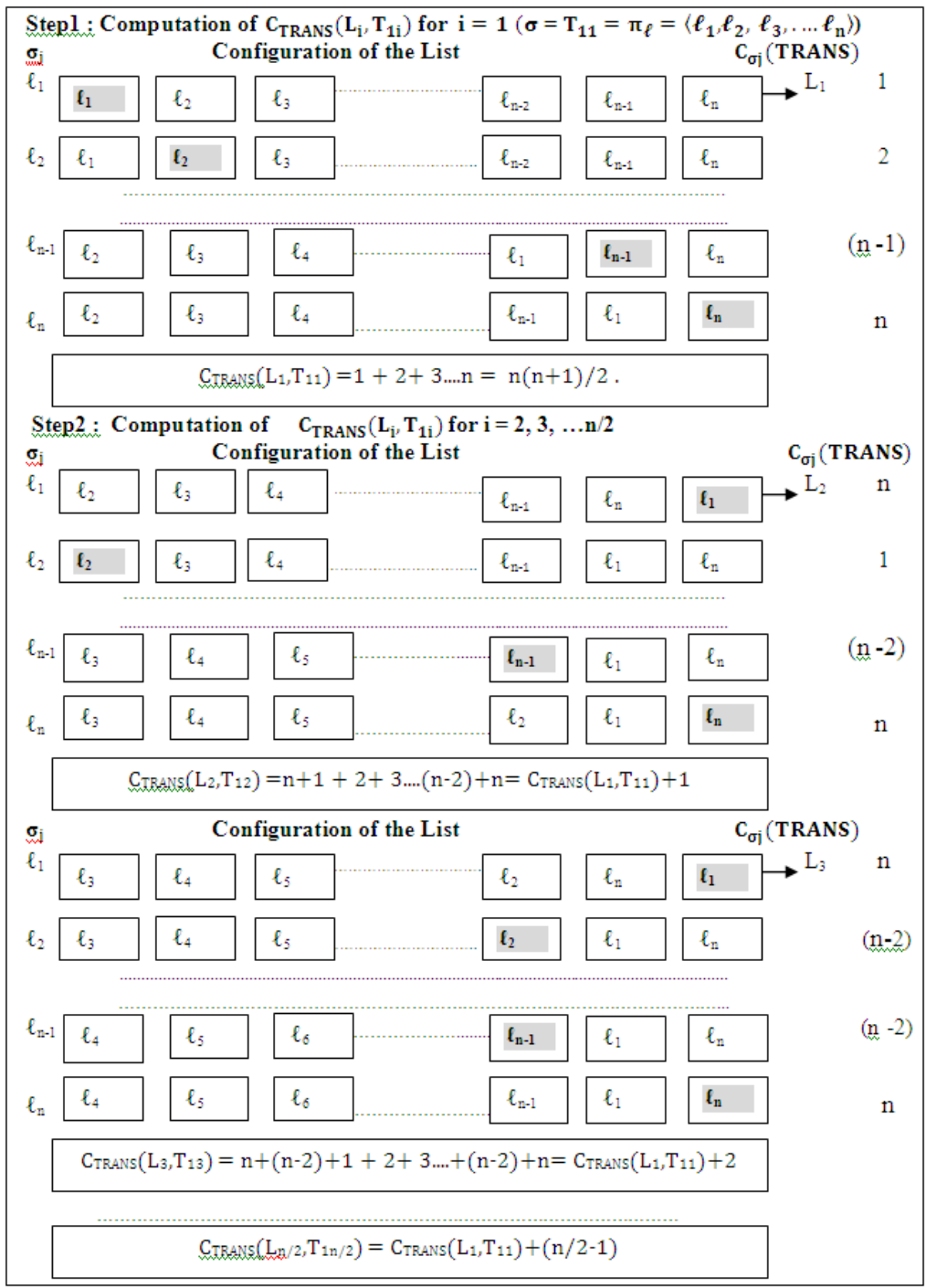

**Figure 1(n is even and k ≤ n/2)**

---

**Theorem 3.2**- *Let* $C_{TRANS}(\ell, T_2)$ *be the total access cost incurred by TRANS algorithm while serving a request sequence of Type* $T_2$, *on a list* $\ell$ *of size n where* $T_2 = (\Pi_r)^k$ *for* $k \geq 1$ *then,*

a) $C_{TRANS}(\ell, T_2) = k \times \left(\frac{n^2+2n}{2}\right)$ *when n is an even number.*

b) $C_{TRANS}(\ell, T_2) = (k) \times \left[\frac{n^2+2n-3}{2} + 1\right]$ *when n is an odd number.*

---

**Proof:** Let $C_{TRANS}(\ell, T_2)$ be the total access cost incurred by TRANS algorithm while serving a request sequence of type $T_2$ on a given list $\ell$ with initial list configuration $<\ell_1, \ell_2, \ell_3 \ldots \ell_n>$. Let $T_{2i}$ be a subsequence of $T_2$ for $i=1, 2, 3, \ldots, k$. So $T_2 = T_{21}T_{22}T_{23}\ldots T_{2k}$, where each $T_{2i} = \Pi_r = <\ell_n, \ell_{n-1}\ldots\ell_2, \ell_1>$ for $i=1, 2, 3, \ldots, k$. Let $C_{TRANS}(L_i, T_{2i})$ be the total access cost of serving a request subsequence $T_{2i}$ of $T_2$ on a list configuration $L_i$. Here $L_i$ denotes a configuration of the list $\ell$ before serving the subsequence $T_{2i}$ for $i=1, 2, 3, \ldots, k$. The total access cost $C_{TRANS}(\ell, T_2)$ can be calculated as $C_{TRANS}(\ell, T_2) = \sum_{i=1}^{k} C_{TRANS}(L_i, T_{2i}) = C_{TRANS}(L_1, T_{21}) + C_{TRANS}(L_2, T_{22}) + \ldots + C_{TRANS}(L_k, T_{2k})$

**Step 1: Computation of $C_{TRANS}(L_i, T_{2i})$ for i=1**

Let σ be a type $T_{21}$ request subsequence of $T_2$ that is served with list configuration $L_1 = \ell = <\ell_1, \ell_2, \ell_3 \ldots \ell_n>$. Let $\sigma_j$ be the $j^{th}$ request of the request subsequence σ and $C_{\sigma j}(TRANS)$ denotes the access cost of serving a request $\sigma_j$ for $j=1, 2, 3 \ldots n$ using TRANS algorithm. Hence $C_{TRANS}(L_1, T_{21}) = \sum_{j=1}^{n} C_{\sigma j}(TRANS)$

**Case i)-** Let n is an even number. Then $C_{\sigma 1}(TRANS) = n$, $C_{\sigma 2}(TRANS) = n$, $C_{\sigma 3}(TRANS) = (n-2), \ldots C_{\sigma(n-1)}(TRANS) = 2$, $C_{\sigma n}(TRANS) = 2$. Hence $C_{TRANS}(L_1, T_{21}) = \sum_{j=1}^{n} C_{\sigma j}(TRANS) = n + n + (n-2) + (n-2) \ldots + 2 + 2 = [n^2+2n]/2$.

**Case ii)-** Let n is an odd number. Then $C_{\sigma 1}(TRANS) = n$, $C_{\sigma 2}(TRANS) = n$, $C_{\sigma 3}(TRANS) = (n-2)\ldots C_{\sigma(n-1)}(TRANS) = 3$, $C_{\sigma n}(TRANS) = 1$. Hence $C_{TRANS}(L_1, T_{21}) = \sum_{j=1}^{n} C_{\sigma j}(TRANS) = n + n + (n-2) + (n-2) \ldots + 3 + 3 + 1 = [n^2+2n-3]/2 + 1$.

**Step 2: Computation of $C_{TRANS}(L_i, T_{2i})$ for i=2, 3, ...., k**

**Case i)-** Let n is an even number. Then $C_{TRANS}(L_i, T_{2i}) = n + n + (n-2) + (n-2) \ldots 2 + 2 = [n^2+2n]/2$ for each $i=2, 3, \ldots, k$. Hence $\sum_{i=2}^{k} C_{TRANS}(L_i, T_{2i}) = (k-1) \times \frac{n^2+2n}{2}$

**Case ii)-** Let n is an odd number. Then $C_{TRANS}(L_i, T_{2i}) = n + n + (n-2) + (n-2) \ldots 3 + 3 + 1 = [n^2+2n-3]/2 + 1$ for each $i = 2, 3, \ldots, k$. Hence $\sum_{i=2}^{k} C_{TRANS}(L_i, T_{2i}) = (k-1) \times \left[\frac{n^2+2n-3}{2} + 1\right]$

**Step 3: Computation of $C_{TRANS}(\ell, T_2)$**

**Proof of Theorem 3.2-a):** Let n is even number. The total access cost $C_{TRANS}(\ell, T_2) = C_{TRANS}(L_1, T_{21}) + \sum_{i=2}^{k} C_{TRANS}(L_i, T_{2i}) = \frac{n^2+2n}{2} + (k-1) \times \frac{n^2+2n}{2} = k \times \frac{n^2+2n}{2}$

**Proof of Theorem 3.2-b):** Let n is odd number. The total access cost $C_{TRANS}(\ell, T_2) = C_{TRANS}(L_1, T_{21}) + \sum_{i=2}^{k} C_{TRANS}(L_i, T_{2i}) = \left[\frac{n^2+2n-3}{2} + 1\right] + (k-1) \times \left[\frac{n^2+2n-3}{2} + 1\right] = (k) \times \left[\frac{n^2+2n-3}{2} + 1\right]$

*2.4 Graphical Representation of Results*

We have compared the performance of MTF and TRANS algorithms for T1 and T2 types of request sequences. Our comparison results are graphically shown in figure 2 and figure 3. Let $C1 = C_{MTF}(\ell, T_1)$, $C2 = C_{TRANS}(\ell, T_1)$, $C3 = C_{MTF}(\ell, T_2)$ and $C4 = C_{TRANS}(\ell, T_2)$. In figure 2, keeping the value of n constant we plot a graph by taking values of k in x-axis and total access costs C1 and C2 in y-axis. Similarly in figure 3, keeping the value of n constant we plot a graph by taking values of k in x-axis and total access costs C3 and C4 in y-axis.

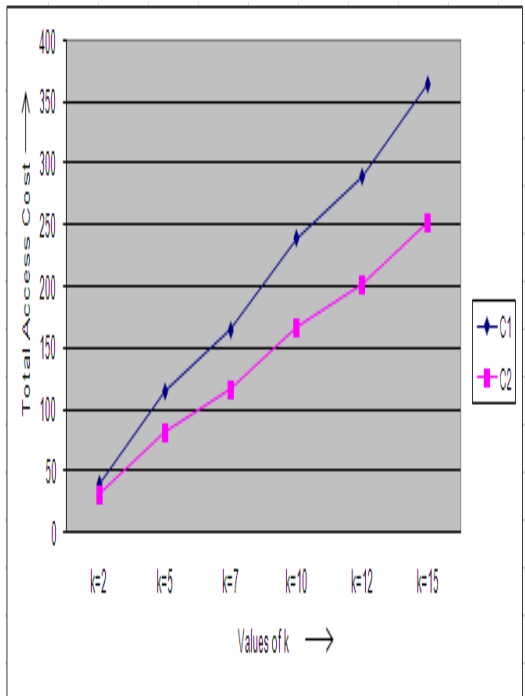
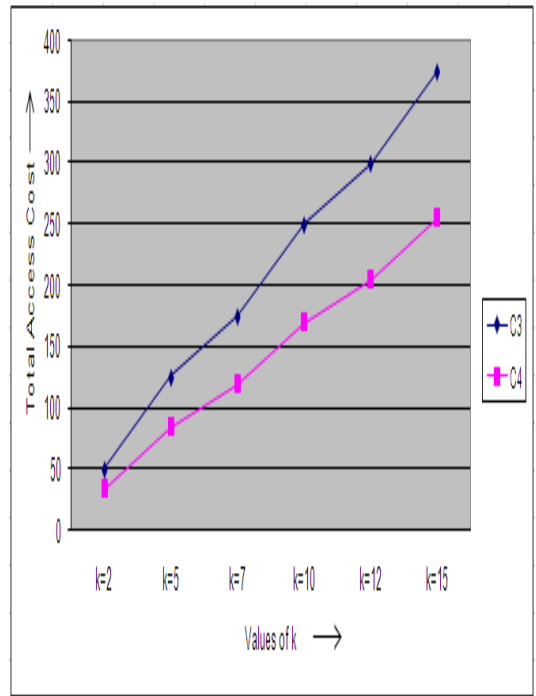

Figure 2  For constant n (n= 5)        Figure 3  For constant n (n = 5)

## 4  Conclusion

In this paper we have generated two different types of request sequences corresponding to real life inputs without locality of reference. Using these request sequences, we have analyzed the performance of TRANS algorithm. We have obtained some novel and interesting theoretical results for computing the total access cost. We have made a comparison of performance of MTF and TRANS list accessing algorithms for these request sequences and represented our comparison results as a graph. Our analytical results show that for two specific types of request sequences without locality of reference TRANS performs better than MTF.

More such types of real life request sequences without locality of reference can be generated and comparative performance evaluation of various list accessing algorithms can be done as a future work.